\documentclass[twocolumn,aps,showpacs,preprintnumbers,amsmath,amssymb, superscriptaddress]{revtex4-1}
\pdfoutput=1

\usepackage{bm}
\usepackage{graphicx}
\usepackage{color}

\begin{document}

\title{Is BaCr$_2$As$_2$ symmetrical to BaFe$_2$As$_2$ with respect to half $3d$ shell filling?}

\author{P. Richard}\email{p.richard@iphy.ac.cn}
\affiliation{Beijing National Laboratory for Condensed Matter Physics, and Institute of Physics, Chinese Academy of Sciences, Beijing 100190, China}
\affiliation{School of Physical Sciences, University of Chinese Academy of Sciences, Beijing 100190, China}
\affiliation{Collaborative Innovation Center of Quantum Matter, Beijing, China}
\author{A. van Roekeghem}
\affiliation{CEA, LITEN, 17 Rue des Martyrs, 38054 Grenoble, France}
\author{B. Q. Lv}
\affiliation{Beijing National Laboratory for Condensed Matter Physics, and Institute of Physics, Chinese Academy of Sciences, Beijing 100190, China}
\affiliation{School of Physical Sciences, University of Chinese Academy of Sciences, Beijing 100190, China}
\author{Tian Qian}
\affiliation{Beijing National Laboratory for Condensed Matter Physics, and Institute of Physics, Chinese Academy of Sciences, Beijing 100190, China}
\affiliation{Collaborative Innovation Center of Quantum Matter, Beijing, China}
\author{T. K. Kim}
\affiliation{Diamond Light Source, Harwell Campus, Didcot, OX11 0DE, United Kingdom}
\author{M. Hoesch}
\affiliation{Diamond Light Source, Harwell Campus, Didcot, OX11 0DE, United Kingdom}
\author{J.-P. Hu}
\affiliation{Beijing National Laboratory for Condensed Matter Physics, and Institute of Physics, Chinese Academy of Sciences, Beijing 100190, China}
\affiliation{School of Physical Sciences, University of Chinese Academy of Sciences, Beijing 100190, China}
\affiliation{Collaborative Innovation Center of Quantum Matter, Beijing, China}
\author{Athena S. Sefat}
\affiliation{Materials Science and Technology Division, Oak Ridge National Laboratory, Oak Ridge, Tennessee 37831-6114, USA}
\author{Silke Biermann}
\affiliation{Centre de Physique Th\'{e}orique, Ecole Polytechnique, CNRS UMR 7644, Universit\'{e} Paris-Saclay, 91128 Palaiseau, France}
\affiliation{Coll\`{e}ge de France, 11 place Marcelin Berthelot, 75005 Paris, France}
\affiliation{European Theoretical Synchrotron Facility, Europe}
\author{H. Ding}
\affiliation{Beijing National Laboratory for Condensed Matter Physics, and Institute of Physics, Chinese Academy of Sciences, Beijing 100190, China}
\affiliation{School of Physical Sciences, University of Chinese Academy of Sciences, Beijing 100190, China}
\affiliation{Collaborative Innovation Center of Quantum Matter, Beijing, China}

\date{\today}

\begin{abstract}
We have performed an angle-resolved photoemission spectroscopy study of BaCr$_2$As$_2$, which has the same crystal structure as BaFe$_2$As$_2$, a parent compound of Fe-based superconductors. We determine the Fermi surface of this material and its band dispersion over 5 eV of binding energy. Very moderate band renormalization (1.35) is observed for only two bands. We attribute this small renormalization to enhanced direct exchange as compared to Fe in BaFe$_2$As$_2$, and to a larger contribution of the $e_g$ orbitals in the composition of the bands forming the Fermi surface, leading to an effective
valence count that is reduced by Fe $d$ - As $p$ hybridization. 
\end{abstract}



\maketitle

\section{Introduction}

The electronic correlations are widely believed to play a major role for unconventional superconductivity in the Fe-based superconductors \cite{AvRoekeghem_CR_Physique17}. It was established theoretically that the strength of the electronic correlations in these materials is tuned by the filling of the $3d$ shell in presence of a large Hund's coupling \cite{HauleNJP11,AichhornPRB82,LiebschPRB82,IshidaPRB81,Georges_ARCMP4, WernerNatPhys8}. Accordingly, angle-resolved photoemission spectroscopy (ARPES) studies report that the band renormalization factor $1/Z$ decreases monotonically upon filling the $3d$ shell: $1/Z=3$ in BaFe$_2$As$_2$ ($d^6$) \cite{RichardPRL2010} (consistent with density functional theory (DFT) + dynamic mean-field theory (DMFT) calculations \cite{Yin_ZP_NMAT10}), $1/Z=1.4$ in BaCo$_2$As$_2$ ($d^7$) \cite{Nan_XuPRX3,Mansart_PRB94}, $1/Z=1.1$ in SrNi$_2$As$_2$ ($d^8$) \cite{ZengPRB94}, and $1/Z=1$ in BaCu$_2$As$_2$ ($d^{10}$) \cite{Wu_PRB91}. Within this framework, the electronic correlations should evolve similarly as a function of band filling with respect to the half $3d$ shell case ($d^5$). Therefore, the $d^4$ case of Cr$^{2+}$, the symmetric counterpart of the $d^6$ filling of Fe$^{2+}$, raises the possibility of unconventional superconductivity for compositions in proximity of BaCr$_2$As$_2$. As with BaFe$_2$As$_2$, BaCr$_2$As$_2$ is an antiferromagnetic metal, and a sizable renormalization factor of 2 was derived from specific heat measurements \cite{Singh_PRB79BaCr2As2}. A recent DMFT study also suggests mass enhancement by a factor of 2 \cite{Edelmann_arxiv}, thus reinforcing the view that BaCr$_2$As$_2$ can be regarded as the symmetrical analog of BaFe$_2$As$_2$. A possible pairing instability is even proposed upon negative pressure or electron doping \cite{Edelmann_arxiv}. However, there is to date no direct experimental characterization of the electronic band dispersion of BaCr$_2$As$_2$ \footnote{After completion of our work, we have been aware of another ARPES study of BaCr$_2$As$_2$ showing results similar to ours \cite{Nayak_arxiv}}, which is essential to answer the question: Is BaCr$_2$As$_2$ symmetrical to BaFe$_2$As$_2$ with respect to half $3d$ shell filling? 

Here we present an ARPES study of BaCr$_2$As$_2$. Despite evidence for a surface state, the Fermi surface (FS) is very similar to the one expected from local density approximation (LDA) calculations. We find that while most bands are not renormalized, two bands near the Fermi energy ($E_F$) are renormalized by a factor 1.35, which is much smaller than reported theoretically. We show that unlike in BaFe$_2$As$_2$, the spectral weight at $E_F$ in BaCr$_2$As$_2$ is significantly contaminated by $e_g$ orbitals and that direct exchange between Fe neighboring atoms play a more significant role. From our experimental results and analysis, we conclude that although there are obvious similarities between BaCr$_2$As$_2$ and BaFe$_2$As$_2$, the analogy between these two compounds cannot be pushed too far. 

\section{Methods}

High-quality single crystals of BaCr$_2$As$_2$ were grown by the self-flux method \cite{Singh_PRB79BaCr2As2}. ARPES measurements were performed using photon energies ($h\nu$) of 56 eV and 73 eV at Beamline I05 of Diamond Light Source equipped with a VG-Scienta R4000 analyzer. The energy and angular resolutions were set at 12 - 15 meV and 0.2$^{\textrm{o}}$, respectively. Additional measurements in the 22 - 80 eV $h\nu$ range have been recorded at the PGM beamline of the Synchrotron Radiation Center equipped with a VG-Scienta R4000 analyzer, with the energy and angular resolutions set at 20 - 30 meV and 0.2$^{\textrm{o}}$, respectively. All samples were cleaved \emph{in situ} and measured at 20 K in a vacuum better than $5\times 10^{-11}$ Torr. Throughout the paper, we label the momentum values with respect to the 1 Cr/unit-cell Brillouin zone (BZ), and use $c^{\prime} = c/2$ as the distance between two Cr planes.

\begin{figure}[!t]
\begin{center}
\includegraphics[width=\columnwidth]{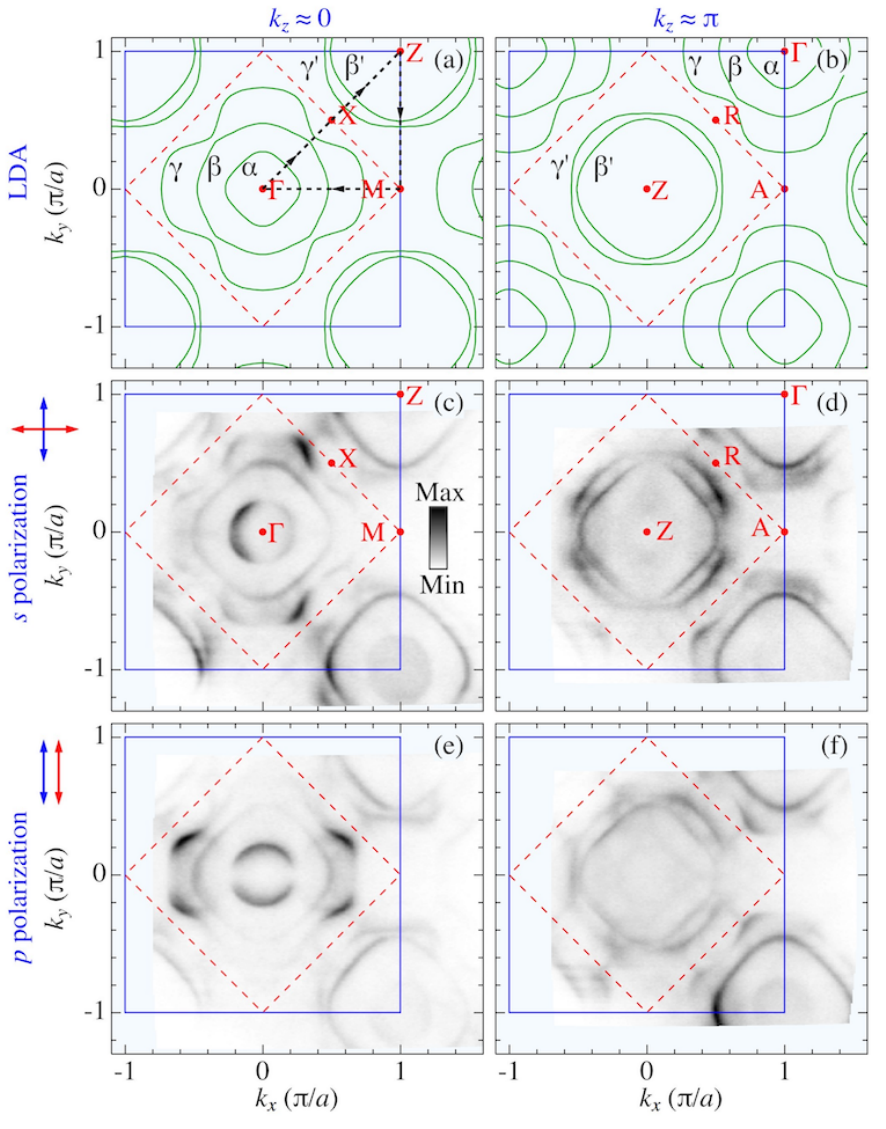}
\end{center}
\caption{\label{FS}(Color online). (a) FS of BaCr$_2$As$_2$ in the antiferromagnetic state calculated within the LDA at $k_z=0$. The blue square indicates the first Brillouin zone (1 Cr/unit-cell notation) boundaries. The black dashed arrows indicate the momentum path of the calculations in Fig. \ref{calc_bands}. (b) Same as (a) but for $k_z=\pi$. (c) ARPES intensity plot integrated within a 10 meV window centered at $E_F$, recorded at 20 K using 73 eV ($k_z\approx 0$), $s$-polarized photons. (d) Same as (c) but with 56 eV ($k_z\approx\pi$) photons. (e) and (f) Same as (c) and (d), respectively, but for $p$-polarized photons. The blue and red arrows on the left side of panels (c) and (e) indicate the direction of the detector slit and of the light polarization, respectively, for the data of the corresponding row. None of the intensity plots has been symmetrized.}
\end{figure}

We have performed DFT calculations within the LDA and in the $G$-type antiferromagnetic order, as implemented in the Wien2k code \cite{Blaha_WIEN2k}.  The lattice parameters were taken from experiment (a = 3.963 \AA, c= 13.600 \AA) \cite{Pfisterer_ZNat35} and we chose z$_{\textrm{As}}$ = 0.3572, similar to what was found in the calculations of Singh \textit {et al.} \cite{Singh_PRB79BaCr2As2}.
 
\section{Results and discussion}

\begin{figure}[!t]
\begin{center}
\includegraphics[width=\columnwidth]{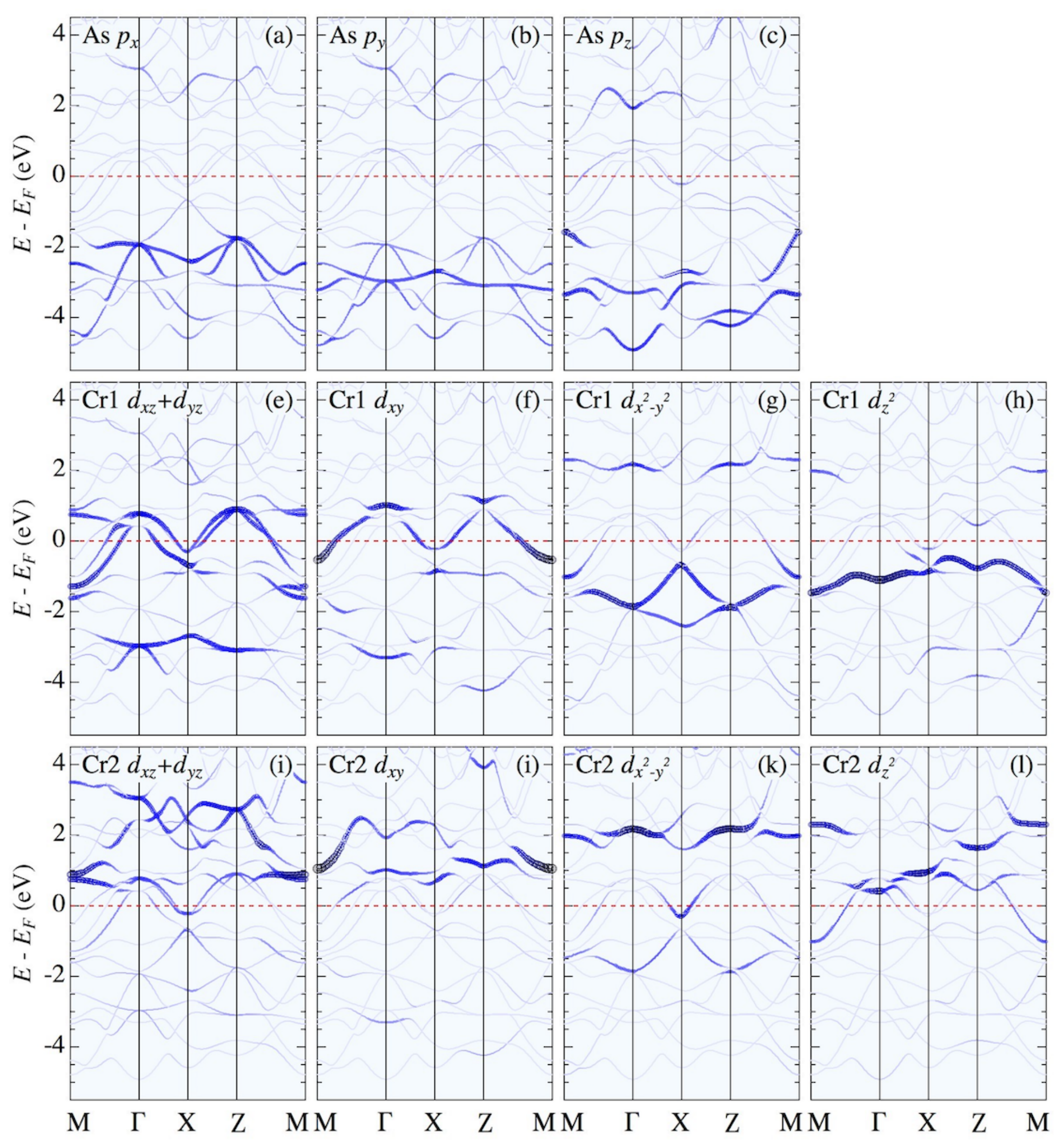}
\end{center}
\caption{\label{calc_bands}(Color online). Orbital projections of our LDA calculations in the AFM state, along the momentum path indicated with dashed arrows in Fig. \ref{FS}(a). The thickness of the lines is proportional to the corresponding orbital content. Cr1 and Cr2 refer to orbitals of Cr with majority and minority spins, respectively.}
\end{figure}

\begin{figure*}[!t]
\begin{center}
\includegraphics[width=\textwidth]{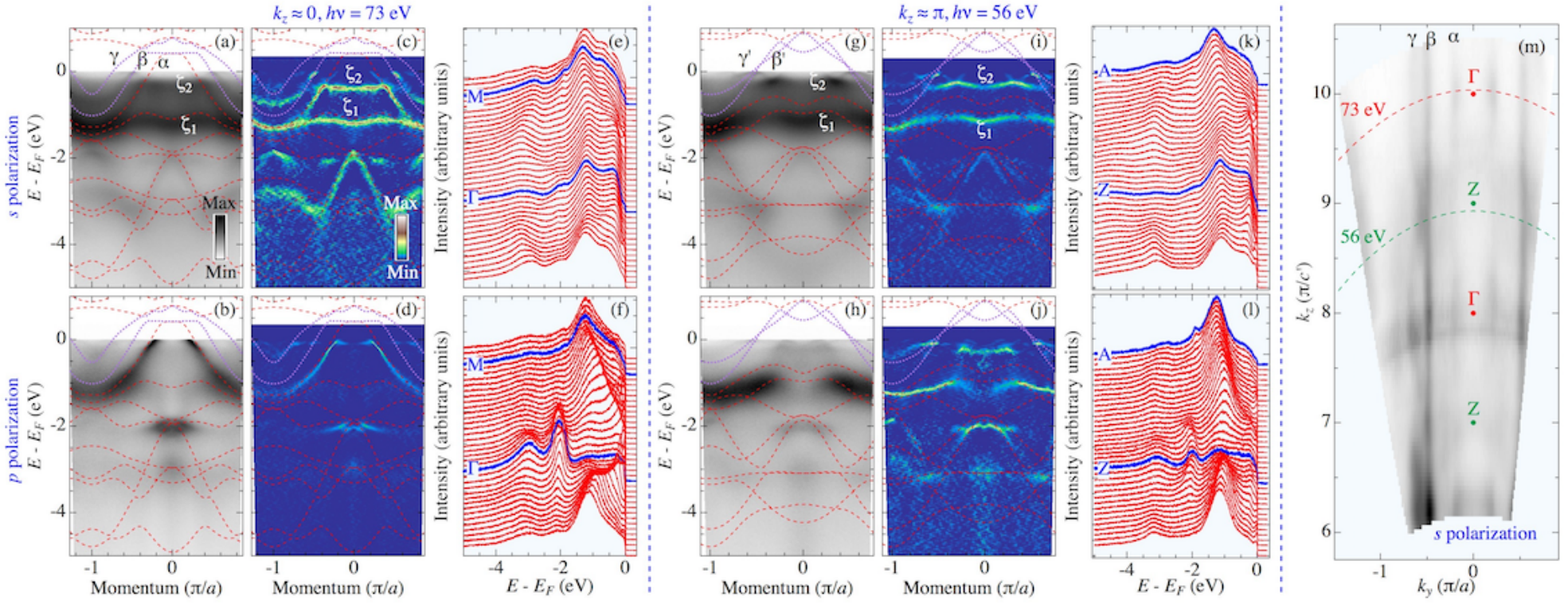}
\end{center}
\caption{\label{dispersion}(Color online). (a) ARPES intensity plot along the $\Gamma$-M direction for $k_z\approx 0$ ($h\nu=73$ eV), recorded with the $s$ polarization configuration. LDA calculations are overlapped for comparison. The line in purple refer to the $\beta$ band. (b) Same as (a) but recorded with the $p$ polarization configuration. (c) Curvature intensity plot corresponding to the data in (a). (d) Curvature intensity plot corresponding to the data in (b). (e) EDCs corresponding to the data in (a). (f) EDCs corresponding to the data in (b). (g)-(l) Same as (a)-(f), but for $k_z\approx\pi$ ($h\nu=56$ eV). (m) Intensity plot of FS in the $k_y-k_z$ plane recorded with $s$-polarized light. The momentum positions of the ARPES cuts recorded at 56 and 73 eV are indicated by dashed lines.} 
\end{figure*}

In Figs. \ref{FS}(a) and \ref{FS}(b), we display our calculations of the FS of BaCr$_2$As$_2$ in the antiferromagnetic state for $k_z=0$ and $k_z=\pi$, respectively. Our results are consistent with previous calculations \cite{Singh_PRB79BaCr2As2,Edelmann_arxiv} and show that the FS differs substantially from that of the Fe-based superconductors. The calculations indicate the existence of 3 hole FS pockets centered at the $\Gamma$ point $(0,0,0)$, and 2 hole FS pockets centered at the Z point $(\pi,\pi,0)=(0,0,\pi)$. Here we call $\alpha$ the inner FS pocket, which is 3D and does not appear around the Z point. The $\beta$ band, which forms the $\beta$ FS in the $k_z=0$ plane, disperses only slightly along $k_z$ and gives the $\beta'$ FS pocket around the Z point. In contrast, the $\gamma$ band varies more strongly along $k_z$ to give a $\gamma$ FS pocket in the $k_z=0$ plane and a $\gamma'$ FS pocket in the $k_z=\pi$ plane that have different shapes and sizes.

The orbital projections of the calculated band dispersions are shown in Fig. \ref{calc_bands}, with the weight of the majority spin (Cr1) and minority spin (Cr2) plotted separately. Our calculations indicate that both the $\alpha$ and $\beta$ FSs derive mainly from Cr1 $d_{xz}+d_{yz}$, but that their orbital composition is not pure and includes other Cr1 and Cr2 $d$ orbitals as well. In contrast, the $\gamma$ band derives almost only from Cr1 $d_{xy}$. Although most of the As $p$ states locate below $E-E_F=-1.5$~eV, our calculations also predict some weight for As $p_z$ at $E_F$. 

The experimental sections of FS displayed in Figs. \ref{FS}(c) and \ref{FS}(e) for $k_z\approx 0$, and in Figs. \ref{FS}(d) and \ref{FS}(f) for $k_z\approx \pi$, are very similar to the calculations. This is notably true for the $\gamma$ and $\gamma'$ FS pockets. Small discrepancies need to be reported though. For example, while the calculations predict a slightly octogonalish shape for the section of the $\beta$ FS around $\Gamma$, what is observed experimentally is more squarish. The $\alpha$ FS in Figs. \ref{FS}(c) and \ref{FS}(e) is also circular, in contrast to the rather squarish calculated one. More importantly, a replica of that FS pocket is detected at the Z point, suggesting a band folding or a surface state. 

Our analysis shows that the $\alpha$, $\beta$ and $\gamma$ sections of FS cover respectively 7.4 \%, 34.5\% and 63.3\% of the 2Cr/unit-cell BZ (dashed red squares in Fig. \ref{FS}). At $k_z\approx \pi$, the $\beta'$ and $\gamma'$ sections of FS cover 50.2 \% and 35.1 \% of the 2Cr/unit-cell BZ. After averaging the total areas at $k_z\approx 0$ and $k_z\approx \pi$, we find 95.3 \% of the 2Cr/unit-cell BZ. Considering spin degeneracy, this coincides to slightly less than the expected $d^4$ filling of Cr$^{2+}$. Our approximation is justified by the $k_y-k_z$ mapping showed in Fig. \ref{dispersion}(m), obtained by scanning $h\nu$ from 22 to 80 eV. The conversion from $h\nu$ to $k_z$ is done using the three-step model with the sudden approximation and the free-electron final state, and we used an inner potential of 13 eV \cite{RichardJPCM27,DamascelliPScrypta2004}. The $\alpha$ FS, small compared to the others, is dispersionless, but has stronger intensity around the $\Gamma$ point, giving further support for the existence of a surface state. While the $\beta$ band is nearly dispersionless along $k_z$, the $\gamma$ band shows sizable dispersion along $k_z$.

In Figs. \ref{dispersion}(a) and \ref{dispersion}(b) we show the electronic band dispersion observed experimentally from ARPES intensity cuts along $\Gamma$-M ($k_x=0$) recorded using 73 eV photons ($k_z\approx 0$) in the $s$ and $p$ polarization configurations, respectively. In a similar fashion, we display in Figs. \ref{dispersion}(g) and \ref{dispersion}(h) the ARPES intensity cuts recorded with 56 eV photons ($k_z\approx \pi$), corresponding to the Z-A high-symmetry line. The LDA calculations are superimposed for comparison. The curvature intensity plots \cite{P_Zhang_RSI2011} and the energy distribution curves (EDCs) corresponding to the ARPES cuts are displayed on their right side. The agreement between experiments and calculations is rather good over 4.5 eV and most bands would need only small energy shifts to match the ARPES intensity plot. While the $s$ polarization configuration favors the observation of states with odd symmetry or with $z$-oriented orbitals such as $d_{z^2}$ or $p_z$, the $p$ polarization configuration favors even symmetry states. Accordingly, the $\gamma$ ($d_{xy}$) band is more intense under $s$ polarization. The experimental results also indicate that the $\beta$ and $\alpha$ bands have dominant odd and even symmetries, respectively.

\begin{figure}[!t]
\begin{center}
\includegraphics[width=\columnwidth]{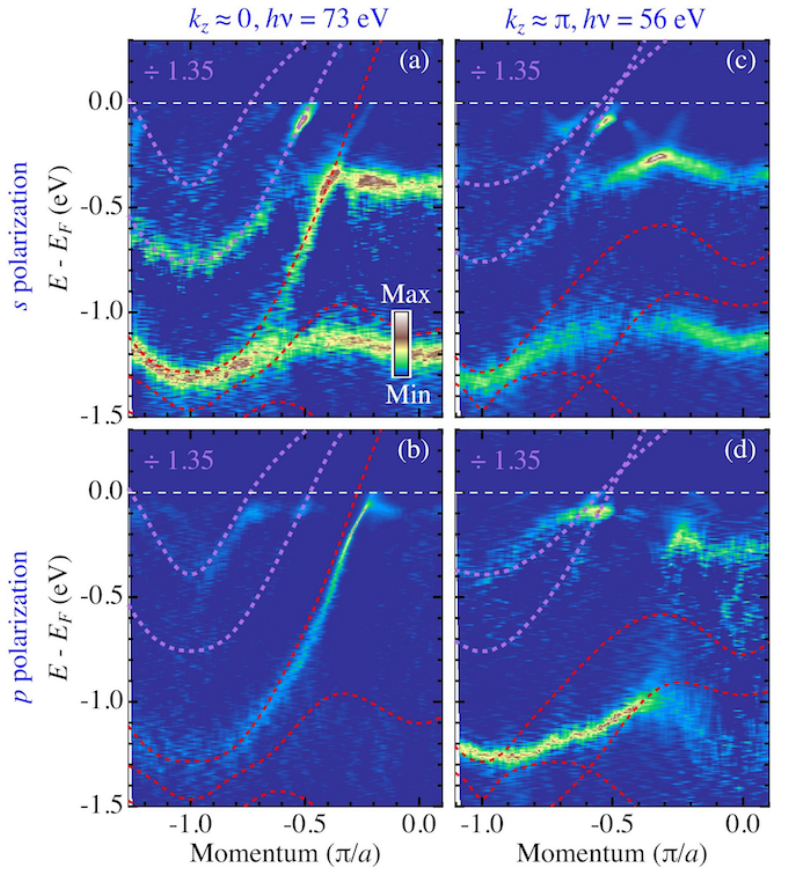}
\end{center}
\caption{\label{Renormalization}(Color online). (a) Near-$E_F$ zoom of the curvature intensity plot obtained with $s$-polarization at $k_z\approx 0$. Dashed lines in red and in purple represent unrenormalized LDA bands and LDA bands renormalized by a factor 1.35, respectively. (b) Same as (a) but recorded with $p$ polarization at $k_z\approx 0$. (c) Same as (a) but recorded with $s$ polarization at $k_z\approx \pi$. (d) Same as (a) but recorded with $p$ polarization at $k_z\approx \pi$.}
\end{figure}

The most obvious difference between the calculations and the experimental results is the presence of the band labeled $\zeta_2$ in Fig. \ref{dispersion}, which is not predicted. As with the $\zeta_1$ band associated to the Cr1 $d_{z^2}$ orbital, it appears clearly only under $s$ polarization. Interestingly, the dispersions of the $\zeta_1$ and $\zeta_2$ bands are identical within uncertainties, and the location of the bottom of the $\zeta_2$ band is nearly the same under both 56 eV and 73 eV photon excitations. Consequently, we conjecture that the $\zeta_2$ band is a surface state with $d_{z^2}$ orbital character. 

As mentioned above, the unrenormalized LDA calculations are in good agreement with the experimental data. Even the $\alpha$ band, near $E_F$, is well reproduced by the unrenormalized LDA calculations. However, the situation is not as good for the $\beta$ and $\gamma$ bands. As we show in Fig. \ref{Renormalization} by zooming on the band dispersion near $E_F$, these two bands fit the calculations better after renormalizing the LDA bands by a factor of 1.35. This factor is far from that observed in the Fe-based superconductors and more consistent with the renormalization found in BaCo$_2$As$_2$ \cite{Nan_XuPRX3,Mansart_PRB94}, which corresponds to a $d^7$ filling. From that perspective, the case of BaCr$_2$As$_2$ is thus not symmetrical to BaFe$_2$As$_2$. 
A similar asymmetry was found theoretically \cite{Pizarro_PRB95} in the 1111 phase (corresponding to LaFeAsO and LaCrAsO). There, it was attributed to the orbital-dependence of the electronic filling \cite{Bascones_PRB86,deMedici_PRL112}. Indeed, the $t_{2g}$ orbitals ($d_{xz}/d_{yz}$ and $d_{xy}$) are closer to half-filling for the $d^6$ configuration (Fe) than for a $d^4$ filling (Cr), whereas the less correlated $e_{2g}$ orbitals ($d_{z^2}$ and $d_{x^2-y^2}$) are the ones closest to half-filling for the $d^4$ configuration \cite{Pizarro_PRB95}. In that picture, recovering a similar electronic mass enhancement as in LaFeAsO would require a $d^n$ filling of around $n=4.6$ to $4.8$.
On the other hand, the situation might not be only determined by the respective orbital fillings.
In fact, another -- possibly more important -- factor is the orbital-resolved kinetic energy that
is larger for orbitals having a low density-of-states at the Fermi level. 
As discussed in Ref. \cite{AichhornPRB82} for FeSe, the orbitals that exhibit a single-particle pseudogap
are less correlated. This can be rationalized by the presence of more possibilities
for fluctuations around the Fermi level in the other three orbitals. A similar mechanism
could be at work in BaCr$_2$As$_2$.

We now try to understand why the 122 structure leads to superconductivity with a high superconducting transition temperature ($T_c$) for Fe-based compounds, whereas no superconductivity is reported for their Cr-based counterparts. The strength of the electronic correlations, which we show in this work to be very different in these two systems, may play an important role. Possible superconductivity was predicted for 3$d^n$ fillings somewhere between $n=4$ and $n=5$ \cite{Pizarro_PRB95,Edelmann_arxiv}, where the strength of the electronic correlations should be comparable to that in Fe-based superconductors. Yet, additional issues need to be addressed. For instance, the magnetic ordering is different in the two systems. BaCr$_2$As$_2$ shows a $G$-type ordering, and a N\'{e}el temperature of 580 K has been reported \cite{Filsinger2017}, with Cr spins pointing along the $c$-axis in EuCr$_2$As$_2$ \cite{Nandi_PRB94}. In contrast, BaFe$_2$As$_2$ orders antiferomagnetically around 140 K with spins aligned inside the Fe plane in a bi-collinear structure \cite{Huang_PRL101}. The antiferromagnetic wave vector in the Fe-based superconductors connects hole and electron pockets FS pockets, giving rise to the quasi-nesting scenario suggesting that low-energy spin fluctuations may contribute to the superconducting pairing \cite{MazinPhysicaC2009, Graser_NJP2009,Cvetkovic_EPL2009}. Although the quasi-nesting scenario is now seriously challenged \cite{RichardJPCM27}, in particular due to the existence of high-$T_c$ superconductivity without hole FS pocket in A$_x$Fe$_{2-x}$Se$_2$ \cite{Qian_PRL2011,XP_WangEPL2011,D_MouPRL2011, Y_Zhang_NatureMat2011} and FeSe monolayers \cite{D_LiuNCOMM2012,S_HeNMAT2013,S_TanNMAT12,JJ_LeeNature515,P_ZhangPRB94_104510}, the $\Gamma$-M wave vectors connects the top of the holelike bands and the bottom of the electron bands within an energy range smaller than a few hundreds of meV in all Fe-based superconductors, suggesting that high-energy spin fluctuations (local moments) may play a role. There is no such electron-hole connection with the antiferromagnetic wave vector in BaCr$_2$As$_2$.  

The difference in the magnetic ordering of BaFe$_2$As$_2$ and BaCr$_2$As$_2$ hides another important property of their respective electronic structures. Although it is known that the exchange interactions between next nearest Fe neighbors (super-exchange $J_2$) is determinant in shaping the magnetic structure of the Fe-based superconductors \cite{J_Zhao_PRL2008,Ewings_PRB78,Diallo_PRL102,McQueeneyPRL2008,J_Zhao_NatPhys2009,LipscombePRL2011}, $G$-type antiferromagnetism is favored by interactions between the nearest Cr neighbors (direct exchange $J_1$). This suggests stronger $d$ orbital mixing in BaCr$_2$As$_2$, which is compatible with our LDA calculations, as mentioned above. It was argued recently that the mixing of the $t_{2g}$ and $e_g$ orbitals (which implies enhanced direct exchange) was detrimental to superconductivity with high $T_c$, and that a suitable electronic configuration that leaves pure and half-filled $t_{2g}$ orbitals at $E_F$ in structures with tetrahedrally-coordinated transition metals was found only for the $d^6$ filling of Fe$^{2+}$ \cite{JP_HuPRX5_041012}. With that respect, BaFe$_2$As$_2$ and BaCr$_2$As$_2$ are very different. Finally, one might speculate that the enhanced $d$-$p$-hybridization leads to an effectively reduced $d$-electron count in BaCr$_2$As$_2$, which would then be consistent
with a correlation strength comparable to the $d^7$ compound BaCo$_2$As$_2$. A similar mechanism was indeed discussed in Ref. \cite{Razzoli_PRB91} where it was highlighted that the correlation strength depends on the effectively available charge rather than on the nominal $d$-electron count.
Finally, one might also speculate that the increased hybridization also influences the effective
crystal field splittings (through ligand field effects) thus modifying more profoundly the scenario of
the Hund's metal phase appearing around the $d^6$ filling. 

\section{Summary}

In summary, we performed an ARPES study of BaCr$_2$As$_2$ that indicates good consistency with LDA calculations in the antiferromagnetic state. A renormalization factor of 1.35 is observed for only two bands, indicating that BaCr$_2$As$_2$ is much less correlated than BaFe$_2$As$_2$. Our analysis suggests a stronger weight of the $e_g$ orbitals at $E_F$, which may be responsible for the observed asymmetry with respect to half $3d$ shell filling. 

\section*{Acknowledgement}

We thank E. Bascones for useful discussions. This work was supported by grants from MOST (Grants Nos. 2015CB921301, 2016YFA0300300 and 2016YFA0401000) and NSFC (Grants Nos. 11274362 and 11674371) from China, a Consolidator Grant of the European Research Council (Project No. 617196) and supercomputing time at IDRIS/GENCI Orsay (Project No. t2016091393). We acknowledge Diamond Light Source for time on beamline I05 under proposal SI9469, which contributed to the results presented here. This work is based in part on research conducted at the Synchrotron Radiation Center, which was primarily funded by the University of Wisconsin-Madison with supplemental support from facility users and the University of Wisconsin-Milwaukee. The work at ORNL was supported by the Department of Energy, Basic Energy Sciences, Materials Sciences and Engineering Division.

\bibliography{biblio_long}

\end{document}